\documentstyle[prd,aps,epsf,eqsecnum]{revtex}
\begin{document}

\title{
Second Order Perturbations in the Randall-Sundrum Infinite Brane-World Model
} 

\author{
Hideaki Kudoh \footnote{ E-mail: kudoh@yukawa.kyoto-u.ac.jp}
and 
Takahiro Tanaka \footnote{ E-mail: tanaka@yukawa.kyoto-u.ac.jp}
}
\address{Yukawa Institute for Theoretical Physics, 
Kyoto University, Kyoto 606-8502, Japan}
\maketitle
\vspace{5mm}

\begin{abstract}
We discuss the non-linear gravitational interactions in the
Randall-Sundrum single brane model. 
If we naively write down the 4-dimensional
effective action integrating over the fifth dimension 
with the aid of 
the decomposition with respect to eigen modes of 4-dimensional 
d'Alembertian, the Kaluza-Klein mode coupling seems to be 
ill-defined. 
We carefully analyze second order perturbations of the gravitational field
induced on the 3-brane under the assumption of the 
static and axial-symmetric 5-dimensional metric. 
It is shown that
there remains no pathological feature in the Kaluza-Klein mode
coupling after the summation over all different mass modes.  
Furthermore, the leading Kaluza-Klein corrections are shown to be 
sufficiently suppressed in comparison with the leading order term 
which is obtained by the zero mode truncation. We confirm that 
the 4-dimensional Einstein gravity is approximately recovered on the
3-brane up to second order perturbations. 

\vspace*{1mm}
\noindent
PACS 04.50.+h; 98.80.Cq~~~~~~~YITP-01-26
\end{abstract}
 
    \section{Introduction}

The possibility that our 4-dimensional universe is embedded as a brane 
in a higher dimensional spacetime has been extensively 
discussed recently as the brane world scenario   
\cite{Arkani-Hamed:1998rs,Antoniadis:1998ig,Randall:1999ee,Randall:1999vf}. 
In particular, Randall and Sundrum introduced attractive models 
whose background bulk spacetime is 5-dimensional anti-de Sitter 
($AdS_5$) spacetime. These models suggest the possibility of the 
existence of extra dimensions in the non-trivial form and 
the possible explanation of the large hierarchy between the 
Planck scale and the electroweak scale\cite{Randall:1999ee,Randall:1999vf}.

The behavior of the gravity in the Randall-Sundrum (RS) models has 
been investigated by many 
authors\cite{Garriga:2000yh,Shiromizu:2000wj,Sasaki:2000mi,Gen:2000nu,Tanaka:2000er,Giddings:2000mu,Tanaka:2000zv,Csaki:2000mp,Charmousis:2000rg,Chiba:2000rr}.  
For the RS single brane model, which is the model with 
the positive tension brane alone, the extension of extra 
dimension is infinite. Nevertheless, 
the gravity on the 3-brane generated by the matter field, which 
is confined on the 3-brane, approximately coincides with the 
4-dimensional Einstein gravity\cite{Garriga:2000yh,Shiromizu:2000wj,Sasaki:2000mi,Gen:2000nu}.

For the RS two-brane model whose bulk space is bounded by the 
positive and the negative tension branes, the 4-dimensional Einstein 
gravity can be  recovered on both branes under the approximation of 
the zero mode truncation\cite{Tanaka:2000er} when we take the 
stabilization mechanism of the distance between the two branes into 
consideration\cite{Goldberger:1999uk,Goldberger:2000un,DeWolfe:2000cp} .

Although a large number of studies have been made on linear 
perturbations of the metric\cite{Kodama:2000fa,Mukohyama:2000ui,Maartens:2000fg,Koyama:2000cc,vandeBruck:2000ju,Langlois:2000ia}, little is known about the non-linear or non-perturbative feature of the gravity 
\cite{Chamblin:2000cj,Chamblin:2000by,Gregory:1993vy,Gregory:2000gf,Chamblin:2001ra,Emparan:2000wa,Dadhich:2000am}.
There is a worry about the RS single brane model. In this model,  
it seems that the non-linear gravitational interaction between Kaluza-Klein (KK) modes becomes strong and it diverges as we move far away from the brane. A more precise argument is that 
the 4-dimensional effective action including interaction terms diverges when we attempt to write it down by integrating over the dependence on the 5th direction by using the decomposition of metric perturbations in terms of eigen modes of 4-dimensional d'Alembertian
\cite{Randall:1999vf}. This fact indicates  that we cannot construct a 4-dimensional effective action for this model in an usual sense.

The above discussion is based on the analysis using the decomposition of the mass eigen modes, 
which is referred to as the mode-by-mode analysis.  
However, it was demonstrated that the mode-by-mode
analysis is insufficient to deal with metric perturbations and it
is necessary to take into account the contributions from all the KK modes
when we discuss the regularity of linear metric perturbations 
at a point far from the 3-brane\cite{Tanaka:2000zv}. 
Thus the pathological behavior in the non-linear
interaction can be expected to be an artifact solely due to the mode-by-mode
analysis, although it has not been proved yet. 
Even if the pathological behavior might be fictitious, 
there is another question if  
the 4-dimensional Einstein
gravity is recovered when we proceed to higher order perturbations?

The purpose of this paper is to give a partial answer to 
the above two questions. 
To study the non-linear behavior of the gravity, 
we investigate second order perturbations 
in the context of the RS single brane model. 
To simplify the analysis, we consider the static and  
axisymmetric configuration, which means that the 3-brane metric is spherically symmetric. Following the method developed by Garriga and Tanaka\cite{Garriga:2000yh}, 
we confirm that there is  
no pathological feature in the KK mode coupling if we sum up 
all the mass eigen modes, and that 
the contribution due to the KK mode coupling is sufficiently small 
compared with the leading order terms, which  
are obtained by the zero mode truncation. 
Furthermore, the result obtained by the zero mode truncation 
exactly agrees with the one predicted by the 4-dimensional Einstein 
gravity.
 \footnote{
  After we had submitted the previous version of this paper, 
  we had the existence of Ref.\cite{Giannakis:2001zx} pointed out. 
  In this reference, the second order metric perturbation in the 
  region far from the star was investigated using the truncation 
  of the first order metric perturbation at the leading order 
  for the expansion with respect to the distance from the star. 
  Since the Einstein equation was not solved in the whole 
  region of the spacetime in their treatment, there remains an 
  ambiguity to add some metric perturbations which satisfy the 
  homogeneous linear perturbation equation only in the asymptotic region.  
  Nevertheless, one may be able to prove that the dominant 
  part of this remaining ambiguity can be absorbed by the redefinition of the mass parameter, and actually our results show that this is the case.  Hence, the recovery of the 4-d Einstein gravity in 
 the region far from the star is to be credited to Ref.\cite{Giannakis:2001zx}. 
  On the other hand, what we show in the present paper is the 
  recovery of the 4-d Einstein gravity throughout the whole spacetime including the inside of the star without assuming a specific radial matter distribution.}

This paper is organized as follows. In the next section we first derive
the second order perturbation equations in the 5-dimensional bulk, and discuss
the gauge transformation as well as the boundary condition.  
In Sec. \ref{sec:linear}, after giving brief summary on 
the behavior of the 5-dimensional Green function, which is need to
solve the perturbation equations, 
we review the results for linear perturbations 
to give their explicit expressions in the notation of the present paper. 
In Sec. \ref{sec:second}, we analyze second order perturbations
of the metric induced on the 3-brane. First it is shown that the
4-dimensional Einstein gravity is recovered in the approximation of
the zero mode truncation, and then we prove that the remaining 
contribution due to the Kaluza-Klein mode coupling can be neglected. 
Section \ref{sec:summary} is devoted to summary.

   \section{perturbation equations in the RS model}
   \label{sec:perturbation equations in RS model} 
 
The brane world model proposed by Randall and Sundrum is 
composed of the 5-dimensional $AdS$ space 
\begin{equation}
    ds^2 =g_{ab}dx^a dx^b=a^2(y)\eta_{\mu\nu}dx^\mu dx^\nu +dy^2\,,
\end{equation}
with a single positive tension ($\sigma >0$) 3-brane located at $y=0$. 
Here $a(y)\equiv e^{-|y|/\ell}$ is the warp factor, and  
$\ell$ is the curvature radius of $AdS_5$. 
We have denoted the 4-dimensional Minkowski metric as
$\eta_{\mu\nu}$. 
The cosmological constant on the bulk and the tension of the 3-brane 
are, respectively,  related to the curvature radius $\ell$ 
as $\Lambda=-6 \ell^{-2}$ and as $\sigma = 3/ 4\pi \ell G_5$, 
where $G_5$ is 5-dimensional Newton's constant. 
The relation between the 5-dimensional and the 4-dimensional Newton's constants 
is given by $G_5=\ell G_4$.  
We also use the notation ${\kappa_5} \equiv 8 \pi G_5$. 
Ordinary matter field is supposed to be localized on the brane. 

In this paper, we investigate second order perturbations 
of this model induced by non-relativistic matter on the brane 
whose energy momentum tensor is given by the perfect fluid form
\begin{equation}
 T_{\mu}^{\nu} = {\mathrm{diag}} \{-\rho,P,P,P\}.
 \label{emtensor}
\end{equation} 
To simplify the analysis, 
we restrict our consideration 
to the static and axisymmetric spacetime 
whose axis of symmetry lies along $y$-direction. 
Namely, the 3-brane metric is spherically symmetric.

We denote the perturbed metric by $ \tilde g_{ab}=g_{ab}+h_{ab} $. 
At the level of linear perturbations,
it is  advantageous to use the Randall-Sundrum (RS) gauge defined by
\begin{eqnarray}
     h_{yy} = h_{\mu y}=0, \quad 
    \label{eq:RS gauge1}
\\
    h ^{\nu}{}_{\mu ;\nu}=0,    \quad   
    h^{\mu}{}_{\mu}=0,
    \label{eq:RS gauge2}
\end{eqnarray}
because the linear perturbation 
equations in this gauge take the simple form \cite{Garriga:2000yh}
\begin{equation}
 {\cal{L}} \, h_{\mu \nu} \equiv
    \left[a^{-2}\Box^{(4)} + \partial_y^2 -4\ell^{-2}\right]
    h_{\mu\nu}=0,
    \label{eq:RS equation}
\end{equation}
and all components are decoupled. 
However, when we consider second order perturbations, 
we cannot impose the transverse-traceless condition (\ref{eq:RS gauge2}) in addition to (\ref{eq:RS gauge1}).  
Hence, we need to abandon to impose the condition (\ref{eq:RS gauge2})
in second order perturbations. 
As a consequence, the second order perturbation equations 
are inevitably coupled.

Here, instead of requiring (\ref{eq:RS gauge2}), 
we start with the assumption of the diagonal form of metric 
\begin{eqnarray}
    ds^2 =a^2
    \left[ -e^{{\cal{A}}(r,y)} dt^2 + e^{{\cal{B}}(r,y)} dr^2+ e^{{\cal{C}}(r,y)} r^2 
    (d\theta^2 + \sin^2 \theta d\varphi^2) \right] + dy^2 \,,
\label{eq:metric}
\end{eqnarray}
which is manifestly compatible 
with the gauge condition (\ref{eq:RS gauge1}), 
and does not loose the generality under the 
restriction to the static and axial-symmetric case.  
Furthermore,  we can expect that this assumption  
is also compatible with the condition (\ref{eq:RS gauge2}) 
at the linear order according to  the result obtained in Ref.\cite{Garriga:2000yh}. For the assumed metric form (\ref{eq:metric}), 
the conditions (\ref{eq:RS gauge2}) at the linear level, 
respectively, become
\begin{eqnarray}
    & & {\cal{A}}^{(1)}+{\cal{B}}^{(1)}+2{\cal{C}}^{(1)}=0 \,,
\label{eq:RS gauge - explicit 1}
\end{eqnarray}
and
\begin{eqnarray}
    & &  \partial_r (r^2 {\cal{B}}^{(1)}) - {2 r} {\cal{C}}^{(1)}=0 \,,  
\label{eq:RS gauge - explicit 2}
\end{eqnarray}
where we have expanded ${\cal{A}}, {\cal{B}}$ and ${\cal{C}}$ to the second order like 
\begin{eqnarray}
    {\cal{A}} (r,y) = \sum_{J=1,2} {\cal{A}}^{(J)} (r,y) \,.  
\label{eq:def A^(J)}
\end{eqnarray}
Hereafter, we neglect higher order terms without mentioning it.

Before we start to solve the 5-dimensional Einstein equation, 
we would like to mention the boundary condition at $y\to\infty$. 
An important point which we must mention here 
is that we are to find a solution which is 
regular at $y\to\infty$. 
If we allow the violation of regularity at the infinity, 
the dynamics of 
the RS brane world model is not uniquely determined.  
Then, such a model cannot be a candidate of the model that 
describes our universe. 
Hence, we require that the metric 
converges to the $AdS_5$ asymptotically.  To guarantee this asymptotic condition, we simply require that the metric functions ${\cal{A}}, {\cal{B}}$ and ${\cal{C}}$ to go to 0 at $y\to\infty$. With this choice, the residual gauge degrees of freedom are completely fixed. Under this constraint, it is still possible to extend the coordinates throughout the bulk maintaining the diagonal form of the metric. We extensively refer to this gauge choice as the RS gauge because we see later that the first order quantities ${\cal{A}}^{(1)}, {\cal{B}}^{(1)}$ and ${\cal{C}}^{(1)}$ in this gauge satisfy the both conditions (\ref{eq:RS gauge - explicit 1}) and (\ref{eq:RS gauge - explicit 2})  (, although the quantities at the second order do not satisfy the transverse-traceless condition (\ref{eq:RS gauge2})).

\subsection{5D Einstein Equations in the bulk}

We consider the 5-dimensional Einstein equations and derive the perturbation equations up to the second order. Since the trace of metric perturbations vanishes in the RS gauge at the linear order, it would be convenient to introduce
\begin{equation}
    2 {\psi}^{(J)} \equiv {\cal{A}}^{(J)}  +{\cal{B}}^{(J)}  + 2 {\cal{C}}^{(J)} \,.
    \label{eq:definition of phi}
\end{equation}
By using this quantity, the trace of metric perturbations is expressed as
\begin{eqnarray}
   \tilde{{g}}^{\mu\nu} {h}_{\mu \nu} 
   = 2  \sum _{J=1,2} {\psi}^{(J)}  - \frac{1}{2} 
    \left[  ({\cal{A}}^{(1)})^2  + ({\cal{B}}^{(1)}) ^2 + 2 ({\cal{C}}^{(1)}) ^2  \right] \,.
\end{eqnarray}
Hence the traceless condition at the first order is simply given by $\psi^{(1)}=0$.

The 5-dimensional vacuum Einstein equation with the cosmological term is equivalent to the following set of equations for the Ricci tensor,
\begin{eqnarray}
   { {R}^y_y} +\frac{4}{\ell^2} &=&  \sum _{J=1,2} \left( \frac{2}
  {\ell} {\psi}^{(J)}_{,y} -{\psi}^{(J)}_{,yy}\right) 
  - Q_{yy} =0    \,,
\label{eq:Einstein eq (yy)}
 \\
     { {R}^t_t} +\frac{4}{\ell^2}   
    &=& \frac{1}{2\ell} \sum _{J=1,2} \left( 2 {\psi}^{(J)}_{,y}  + 4 {\cal{A}}^{(J)}_{,y} 
    - \ell {\cal{A}}^{(J)}_{,yy}  - \ell a^{-2} \triangle {\cal{A}}^{(J)} \right) 
    - \frac{1}{2a^2} {S} =0 \,,
\label{eq:Einstein eq (tt)}
\\
  {R}^y_r&=& \frac{1}{2r}
 \sum_{J=1,2} 
 \Bigl[
   \frac{1}{r^2} (r^3 {\cal{B}}^{(J)})_{,ry}-2(r{\psi}^{(J)}_{,y})_{,r} 
   + {\cal{A}}^{(J)}_{,y}  
 \Bigr]
 +\frac{1}{2} \psi ^{(1)}_{,r} {\cal{B}}^{(1)}_{,y} - Q_{ry} =0  ,
\label{eq:Einstein eq (yr)}
\end{eqnarray}
where $\triangle\equiv \sum_{i=1}^3 \partial_i^2$ and we have defined 
\begin{eqnarray}
    S(r,y) &\equiv& - \frac{1}{r^2} \partial_r 
    \left( r^2 {\cal{A}}^{(1)} _{,r} {\cal{B}}^{(1)}  \right)
    + a^{2} {\psi}^{(1)}_{,y} {\cal{A}}^{(1)} _{,y} + {\psi} ^{(1)}_{,r} {\cal{A}}^{(1)} _{,r}  ~\,,
\label{defS}
\\
Q_{\mu\nu} (r,y) &\equiv& 
\frac{1}{4} 
\left( {\cal{A}}^{(1)}_{,\mu} {\cal{A}}^{(1)}_{,\nu} 
  + {\cal{B}}^{(1)} _{,\mu} {\cal{B}}^{(1)} _{,\nu} 
  + 2 {\cal{C}}^{(1)}_{,\mu} {\cal{C}}^{(1)} _{,\nu}
   \right).
\end{eqnarray}
Note that no other equations for the remaining components 
are independent.

First we consider the trace part ${\psi}^{(J)}$, which can be evaluated by  integrating (\ref{eq:Einstein eq (yy)}). From the requirement of the boundary condition, ${\psi}^{(J)}$ must go to $0$ at $y\to\infty$.  
Thus, we obtain 
\begin{eqnarray}
   {\psi} ^{(J)}(r,y) =
    -\epsilon^{(J)} \int^y_{\infty} {dy''}{e^{2y''/\ell}}
    \int^{y''} _{\infty} dy' e^{-2y' /\ell} Q_{yy} ~. 
\label{eq:phieq}
\end{eqnarray}
where we have introduced a symbol 
$\epsilon^{(J)}$ that is defined by
$\epsilon^{(1)}=0$ and  $\epsilon^{(2)}=1$ to represent 
the first and the second order equations in 
a single expression.  
As is anticipated above, we can see from Eq.~(\ref{eq:phieq}) that 
the traceless condition at the linear order is actually satisfied, while  
that at the second order cannot be imposed any longer in general. 

Let us now turn to ${\cal{B}}^{(J)}$.  
Integrating (\ref{eq:Einstein eq (yr)}), we obtain 
\begin{eqnarray}
  {\cal{B}} ^{(J)}(r,y) 
   = - \frac{1}{r^{3}} \int  r^{2}{\cal{A}} ^{(J)} dr
      + \frac{2}{r^{3}} \, \epsilon^{(J)}   
      \left( \int dr \, r^{2}( r {\psi}^{(2)} )_{, r} 
  +   \int_{\infty}^y dy 
      \int  dr \ r^{3}  Q_{ry} 
      \right) \,. 
    \label{eq:solved B^(J) in GN gauge}
\end{eqnarray}
As before, the integration constant is fixed by the boundary condition at $y\to\infty$.  Now, combined with Eq. (\ref{eq:RS gauge - explicit 1}),  it is easy to see that Eq.(\ref{eq:RS gauge - explicit 2}) holds at the linear order. Hence, it is confirmed that our choice of gauge is equivalent to the RS gauge.

So far, we obtained the relations between the metric functions 
${\cal{A}}^{(J)}$, ${\cal{B}}^{(J)}$ and ${\cal{C}}^{(J)}$. 
Substituting these into Eq.~(\ref{eq:Einstein eq (tt)}), 
we obtain a single equation for ${\cal{A}}^{(J)}$, 
\begin{eqnarray} 
  {\cal{L}}  \bigl[ a^2 {\cal{A}}^{(J)} \bigr]
  =  \epsilon^{(J)} \bigl[ 2a^2 \ell^{-1} \psi_{,y} -S \bigr] \,. 
\label{master}
\end{eqnarray}
The source terms are absent at the linear order,  as they are composed of the quadratic in the linear order quantities.  Since $\psi^{(1)}=0$, 
$S$ defined in Eq.(\ref{defS}) simplifies as
\begin{eqnarray}
    S(r,y) = \frac{1}{r^2} \partial_r \left( r^2 {\cal{A}}^{(1)} _{,r}  {\cal{B}}^{(1)} \right) . 
\label{eq:S_1}
\end{eqnarray}

\subsection{Boundary condition}

In the previous subsection, we have obtained the master equation 
for the metric functions in the bulk up to the second order. 
To solve this equation, we need to know the boundary condition 
to be imposed on the 3-brane. 
The boundary condition on the 3-brane is 
specified by the Israel's junction condition\cite{Israel:1966}. 
However, in the RS gauge defined above, the $y$-constant surface 
is, by construction, chosen so that the metric functions 
go to 0 at $y\to \infty$. 
After fixing the $y$-constant surface for large $y$, 
coordinates are extended to the region near the 3-brane.   
Therefore, we can no longer expect that 
the location of the 3-brane coincides with the $y=0$ surface 
in general\cite{Garriga:2000yh}. In such coordinates, the junction condition is not so trivial. Thus it will be  convenient to introduce other coordinates $\bar{x}^a$, in which the location of the 3-brane stays at $\bar y=0$ but the metric form is still kept diagonal. 
We associate a bar with the quantities written in these coordinates, 
say, ${\bar{\cal{A}}}$. We denote this choice of coordinates by the Gaussian normal (GN) gauge. By construction, $\bar h_{\mu\nu}|_{\bar y=0}$ gives the 4-metric induced on the 3-brane.

The junction condition on the 3-brane is simply written in the GN gauge as
\footnote{
One may think that 
this equation should have been expressed as  
\begin{eqnarray*}
    \tilde{\quad \bar{g} ^{\gamma\nu}}(\bar x) 
        \partial_{\bar y} (a^{-2}(\bar y)\bar h_{\mu\nu}(\bar x)) 
    = - {\kappa_5}  
    \left( T^\gamma{}_\mu - {1 \over 3} 
    \delta^\gamma{}_\mu \, T^\lambda{}_\lambda  \right)(\bar {\bf x}).   
    \qquad  (\mbox{at}\quad \bar y=0+), 
\end{eqnarray*}
because this is a relation for the quantities in the GN gauge. 
However, $f(x)|_{y=0}=0$ is obviously equivalent to 
$f(\bar x)|_{\bar y=0}=0$ for any function $f$. 
To stress this point, we always use the coordinates without bar hereafter. The only exception is Eq.~(\ref{gaugeT}). }
\begin{equation}
    \tilde{\quad \bar{g} ^{\gamma\nu}}(x) 
        \partial_{y} (a^{-2}(y)\bar h_{\mu\nu}(x)) 
    = - {\kappa_5}  
    \left( T^\gamma{}_\mu - {1 \over 3} 
    \delta^\gamma{}_\mu \, T^\lambda{}_\lambda  \right)({\bf x}).   
    \qquad  (\mbox{at}\quad y=0+)
    \label{eq:Jsrael's junction condition}
\end{equation}
As mentioned earlier, 
we assume the energy momentum tensor 
of the perfect fluid form (\ref{emtensor}). 
Then, the 4-dimensional energy-momentum conservation $T ^\nu{}_{\mu;\nu}=0$ becomes 
\begin{eqnarray}
 (\rho(r)+P(r))\partial_r {\bar{\cal{A}}}^{(1)}(r,0) 
       + 2 \, \partial_r P(r) =0  \,,
\label{eq:energy momentum cons}
\end{eqnarray}
and hence we find that $P(r)$ is a second order quantity. 
This equation represents the force balance between pressure and 
gravity acting on the matter field.

Taking it into account that $P(r)$ is second order and  
expanding $\rho$ as $\rho=\rho^{(1)}+\rho^{(2)}$, 
the explicit junction conditions for the metric functions become 
\begin{eqnarray}
    \partial _{y} {{\bar{\cal{A}}}^{(J)}} 
  &=& {\kappa_5} \left( \frac{2}{3} \rho^{(J)} 
          + \epsilon^{(J)} P \right) \,,
\nonumber
\\
    \partial _{y} {{\bar{\cal{B}} }^{(J)}} 
  &=&  \partial _{y} {{\bar{\cal{C}} }^{(J)}} 
   = \, - \frac{ {\kappa_5}}{3} \rho^{(J)}  
         \,, \quad  (\mbox{at}\quad y=0+) \,.
\label{eq:Junction condition -explicit form}
\end{eqnarray}

The boundary condition obtained above is written in terms of the 
variables in the GN gauge. 
To interpret the conditions 
(\ref{eq:Junction condition -explicit form}) in terms of the 
variables in the RS gauge, we consider the gauge transformation 
between these two gauges, 
which is defined by 
\begin{equation}
\tilde {\bar g}_{ab}(\bar x)\, d\bar x^a d\bar x^b
 = \tilde {g}_{ab}(x)\, dx^a dx^b,  
\label{gaugeT}
\end{equation}
with $\bar x^a=x^a+\xi^a(x)$. Since $h_{y\mu}$ and $h_{y y}$ vanish in both gauges, the infinitesimal gauge transformation 
$
\xi^a = \stackrel{\scriptscriptstyle(1)}{\xi^a} + \stackrel{\scriptscriptstyle(2)}{\xi^a}
$ 
between them is restricted to the following form 
\begin{eqnarray}
  \stackrel{\scriptscriptstyle(J) \ \qquad }{\xi^y (r,y)} 
   &=&  \stackrel{\scriptscriptstyle(J) \quad \ }{\hat{\xi}^y (r)} 
   - \epsilon^{(J)}  \biggl[ ~
   \frac{\ell}{4a^2}(\stackrel{\scriptscriptstyle(1) \ }{\hat{\xi}^y_{,r}})^2 
   \biggr] \, ,
\label{eq:solved xi^(J)}
\\
  \stackrel{\scriptscriptstyle(J) \ \qquad }{\xi^r (r,y)} 
  &=&  \stackrel{\scriptscriptstyle(J) \quad \ }{\hat{\xi}^r (r)}
   - \frac{\ell}{2a^2} 
     \stackrel{\scriptscriptstyle(J)\ }{\hat{\xi}^y _{, r} }
 + \epsilon^{(J)} 
   \stackrel{\scriptscriptstyle(1) \ }{\hat{\xi}^y_{, r}}
   \biggl[  
  \frac{\ell}{2a^2}
  \bigl(  \stackrel{\scriptscriptstyle(1) \ }{\hat{\xi}^r_{, r}} 
    - \frac{2}{\ell}  \stackrel{\scriptscriptstyle(1) \ }{\hat{\xi}^y}
  \bigr)
    + \int^y_0  \frac{dy'}{a^2} 
     {\bar{\cal{B}}^{(1)} }(r,y')  
  \biggr] \, .
\label{eq:solved xi^r(J)}
\end{eqnarray}
Functions  
$\stackrel{\scriptscriptstyle(J) \quad \ }{\hat{\xi}^y (r)}$ and 
$\stackrel{\scriptscriptstyle(J) \quad \
}{\hat{\xi}^r (r)}$ are integration constants 
which appear as a result of $y$-integration. 
We discuss how these functions are determined soon later.

We denote the difference between the metric in the RS gauge 
and that in the GN gauge as 
\begin{equation}
        \delta {\cal{A}}^{(J)}(r,y) \equiv {\cal{A}}^{(J)}(r,y) - 
       {\bar{\cal{A}}}^{(J)}(r,y) \,.
\label{eq:def -- sA-bsA}
\end{equation}
The remaining metric functions 
$\delta {\cal{B}}^{(J)}$ and $\delta {\cal{C}}^{(J)}$ are defined in the 
same way. Then, the gauge transformations are given by  
\begin{eqnarray}
    {\delta {\cal{A}}}^{(J)}  &=&
     - \frac{2}{\ell}\stackrel{\scriptscriptstyle(J)\ }{\xi^y}
     + \epsilon^{(J)}  \, \Bigl[  {\bar{\cal{A}}}^{(1)} _{,y} 
      \stackrel{\scriptscriptstyle(1)\ }{ \xi^y}
    + \, {\bar{\cal{A}}}^{(1)} _{, r} 
       \stackrel{\scriptscriptstyle(1)\ }{\xi^r}  \Bigr]
   \,,
\nonumber
\\
   \delta {\cal{B}} ^{(J)} &=& 
    - \frac{2}{\ell} \stackrel{\scriptscriptstyle(J)\ }{\xi^y} 
    + 2\stackrel{\scriptscriptstyle(J)~ }{\xi ^r_{, r} }
    + ~  \epsilon^{(J)}  \, 
    \Bigl[ ~ {\bar{\cal{B}}^{(1)} } _{, y}  
       \stackrel{\scriptscriptstyle(1)\ }{\xi^y} 
    + \,  {\bar{\cal{B}}^{(1)} } _{,r}  
       \stackrel{\scriptscriptstyle(1)\ }{\xi^r} 
    - \Bigl( \stackrel{\scriptscriptstyle(1) ~}{\xi ^{r} _{,r}} \Bigr)  ^{2} 
    + e^{2y/\ell} \Bigl( 
        \stackrel{\scriptscriptstyle(1) ~}{\xi^y_{,r}} \Bigr)^{2}
    ~  \Bigr]  \,,
    \label{eq:gauge trans. - dA(J) dB(J) dC(J)}
\\
   \delta {\cal{C}} ^{(J)} &=& 
   - \frac{2}{\ell}\stackrel{\scriptscriptstyle(J)\ }{\xi ^y} 
    + \frac{2}{r} \stackrel{\scriptscriptstyle(J)\ }{\xi ^r}
    + \epsilon^{(J)} \,  \Bigl[ ~ {\bar{\cal{C}}^{(1)} } _{, y} 
          \stackrel{\scriptscriptstyle(1)\ }{\xi^y} 
    + \, {\bar{\cal{C}}^{(1)} } _{, r} 
          \stackrel{\scriptscriptstyle(1)\ }{\xi^r} 
    - \Bigl({1\over r}\!\stackrel{\scriptscriptstyle(1)\ }{\xi^r} \Bigr)^2 
    ~ \Bigr] \,.
\nonumber
\end{eqnarray}

Now we are ready to derive the equation that determines 
$\hat\xi^y(r)$. We evaluate the identity 
$\psi_{,y}=\bar\psi_{,y}+\delta\psi_{,y}$ at $y=0$. 
The expression for the left hand side is obtained by means of   
Eq.(\ref{eq:phieq}), 
while the right hand side is evaluated  
by substituting Eqs.
(\ref{eq:Junction condition -explicit form}), 
(\ref{eq:solved xi^(J)}), 
(\ref{eq:solved xi^r(J)}) and 
(\ref{eq:gauge trans. - dA(J) dB(J) dC(J)}). 
After tedious but straightforward computation, we obtain 
\begin{eqnarray}
    \triangle \! \stackrel{\scriptscriptstyle(J) ~}{\hat{\xi}^y}  &=&
   \frac{{\kappa_5}}{6}  T^{(J)} 
   + \epsilon^{(J)} \Bigl[ \Xi(r) - \psi_{,y}|_{y=0} \Bigr], 
\label{eq:xi^y^(J) and f^(J)}
\end{eqnarray}
with
\begin{eqnarray}
\Xi (r) &\equiv& 
\triangle   \Bigl[ 
  \frac{1}{\ell}\bigl( \stackrel{\scriptscriptstyle(1)~}{\hat{\xi}^y} \bigr)^2 
  + \int^{r}_{\infty} \stackrel{\scriptscriptstyle(1)~~}
      {\hat{\xi}^y _{,r'}} {\cal{B}}^{(1)}(r',0) dr' 
  \Bigr] 
  - \frac{3}{\ell} \bigl( \stackrel{\scriptscriptstyle(1)~}
      {\hat{\xi}^y_{,r}} \bigr)^2 
 + \stackrel{\scriptscriptstyle(1)\,}{\hat{\xi}^r}  \partial_r
  \triangle 
      \stackrel{\scriptscriptstyle(1)\, }{\hat{\xi}^y} 
 + \frac{ \ell}{2} (\stackrel{\scriptscriptstyle(1)~~}{\hat{\xi}^y_{,rr}})^2   
 + \frac{\ell}{r^2} (\stackrel{\scriptscriptstyle(1)~}{\hat{\xi}^y_{,r}})^2   
\end{eqnarray}

The gauge freedom for the radial coordinate in the GN gauge 
has not been fixed.  
Although the simplest choice might be to take 
$\stackrel{\scriptscriptstyle(J)~}{\xi^r}$ to 
vanish on the brane, for later convenience 
we impose the isotropic gauge condition on the 3-brane,  
${\bar{\cal{B}} }={\bar{\cal{C}} }$ at $\bar y=0$. 
Rewriting this condition by using ${\cal{B}}$ and ${\cal{C}}$ with the 
substitution of (\ref{eq:gauge trans. - dA(J) dB(J) dC(J)}),  
we obtain 
\begin{eqnarray}    
   \stackrel{\scriptscriptstyle(1)\ }{\xi^r} 
    &=& -\frac{r}{4} {\cal{B}}^{(1)} \,,
   \label{eq:isot gauge cond for xi^r 1}
\end{eqnarray}
where we have used (\ref{eq:RS gauge - explicit 2}). 
Here we show the result only for the first order, because   
we do not need the explicit expression for $\stackrel{\scriptscriptstyle(2)\ }{\xi^r}$ in the following discussion.

Once we obtain an explicit expression for 
$\stackrel{\scriptscriptstyle(J) ~}{\hat{\xi}^y}$, 
it is easy to derive the boundary condition 
for the metric functions in the RS gauge from the junction condition 
in the GN gauge (\ref{eq:Junction condition -explicit form}).   
Especially, the boundary condition required to solve the 
master equation (\ref{master}) is deduced by  
substituting the relations obtained above into the 
right hand side of the equation, 
${\cal A}_{,y}={\bar{\cal{A}}}_{,y}+\delta{\cal{A}}_{,y}$. 
Imposing this boundary condition is equivalent to add 
a $\delta$-function source localized on the 3-brane. 
The 5-dimensional master equation (\ref{master}) 
including the boundary condition becomes 
\begin{eqnarray} 
   \left[ {\cal{L}} +4 \ell^{-1}\delta (y) \right] \bigl(a^2 {\cal{A}}^{(J)}\bigr)
    = 2 {\kappa_5} \, \Sigma^{(J)}(r) \, \delta (y)
     + \epsilon^{(J)} \Bigl[ 2 a^2\ell^{-1} \psi_{,y} -S \Bigr] \,,
\label{eq:master eq of delA^(J) in RS - Sigma}
\end{eqnarray}
where
\begin{eqnarray}
   \Sigma ^{(J)}  (r) &\equiv& 
  \kappa_5^{-1} \partial_y {\cal{A}}^{(J)}|_{y=0}  \cr
    &=& \frac{2}{3}\rho^{(J)} 
+\epsilon ^{(J)} \Bigl[ P + \kappa_5^{-1} \partial_y (\delta {\cal{A}}
 ^{(2)})|_{y=0}  \Bigr]\,,
\label{sigma}
\end{eqnarray}
By using the Green function which satisfies
\begin{eqnarray}
    \left[ {\cal{L}} +4 \ell^{-1}\delta (y) \right] 
     G({\mathbf{x}},y;{\mathbf{x}}',y')
      = \delta(y-y') \delta^{3}( {\mathbf{x}} - {\mathbf{x}}' ) \, ,
\end{eqnarray}
the formal solution for the master equation 
(\ref{eq:master eq of delA^(J) in RS - Sigma}) 
is  given by  
\begin{eqnarray}
  a ^2 {{\cal{A}}}^{(J)}(r,y)  &=& 
  2 {\kappa_5} \int dx'^3 G({\bf x},y;{\bf x}',0) ~ \Sigma^{(J)} \cr
\label{eq:formal sol. of sA^J}
  && - ~ 2 \, \epsilon^{(J)} \int d^3x'\int_0^{\infty} dy'\, 
          G ({\bf x},y;{\bf x}',y')\, 
    \left[S(r',y')
    + \frac{2}{ \ell}  
        \int_{\infty}^{y'} a^{2}(y'') Q_{yy}(r',y'') dy'' \right]  \,. 
\end{eqnarray}
The factor 2 in the second term of (\ref{eq:formal sol. of sA^J}) 
comes  from $Z_2$-symmetry.
In \S\ref{sec:linear} we discuss some basic aspects of the 
Green function, and solve the above equation at the linear order. 
In \S\ref{sec:second} 
we extend our analysis to the 
second order. 

    \section{Green function and first order perturbations}
    \label{sec:linear}
    \subsection{Green function}
We need the Green function to evaluate explicitly the formal solution
(\ref{eq:formal sol. of sA^J}). 
In the static case Green function is 
\begin{eqnarray}
        G({\mathbf{x}},y;{\mathbf{x}}',y')  = 
        - \frac{1}{4\pi \ell R}  \left( a^2(y)a^2(y') 
        + \ell \int_{0}^{\infty} u_m(y) u_m(y')  e^{-m R}dm \right) \, ,
    \label{eq:static Green fun.}    
\end{eqnarray} 
where $u_m(y)$ is the mode function, and we have introduced the notation 
\begin{eqnarray*}
    R \equiv |\mathbf{x}-\mathbf{x}'|.  
\end{eqnarray*}

The explicit form of the mode function is given in 
terms of Bessel functions as 
\begin{equation}
    u_m(y) = N_m 
	\bigl[ 	J_1(m\ell)Y_2(m\ell/a)-Y_1(m\ell)J_2(m\ell/a) \bigr], 
\end{equation}
with $N_m = \sqrt{m\ell}/\sqrt{2(J_1(m\ell)^2+Y_1(m\ell)^2)}$.  
It is orthonormalized as 
\begin{eqnarray}
 2 \int _0 ^\infty \frac{dy}{a^{2}}  u_m  u_{m'} &=&  \delta (m-m') \, .
 \label{eq:orthogonality of u_m u_m}
\end{eqnarray}
In particular, setting $m'=0$, we have 
\begin{eqnarray}
 \int^{\infty}_{0} {dy} \  u_m &=&  0 \,  \quad(m\ne 0). 
 \label{eq:orthogonality of u_m and 1}
\end{eqnarray} 

The first term on the right hand side 
of (\ref{eq:static Green fun.}) is the contribution 
from the zero mode whose 4-dimensional mass 
eigenvalue is zero $(m=0)$. We denote this part of Green function by $G_0$. 
The second term corresponds to the propagator due to the Kaluza-Klein states
which have nonzero mass eigenvalues $(m>0)$. 
This term is denoted by $G_K$. Thus, $G=G_0+G_K$.

For large separation $R\gg \ell$, 
the existence of the factor $e^{-mR}$ in the second term in 
(\ref{eq:static Green fun.}), i.e. $G_K$, implies that the 
integral over $m$ is dominated by the contribution from small $m$. 
When the source is on the 3-brane $(y'=0)$, 
we can approximately evaluate $G_K$ by expanding Bessel functions 
taking $m\ell$ as small, but here the Bessel function with the argument 
$m\ell/a$ is to be kept unexpanded because $a$ can be exponentially small. Then, the integration over $m$ can be performed for the 
leading power of $m$ explicitly. The result is given by \cite{Garriga:2000yh}
\begin{equation}    
    G({\bf x},y;{\bf x}',0) \approx  
        - {a^3\over 8\pi \ell} 
    \frac{(2a^2R^2+3\ell^2)}{(a^2R^2+\ell^2)^{3/2}} \,.
\label{asymptotic}
\end{equation}
For a small separation $R\ll \ell$, the Green function is 
dominated by the contribution from modes with large $m$. 
In this limit, the Green function behaves as 
the ordinary 5-dimensional one 
$\approx -1/4\pi^2 [R^2+\ell^2(1-a^{-1})^2]$.

\begin{figure}[htbp]
\begin{center}
 \centerline{
  \epsfxsize = 6cm
  \epsfbox{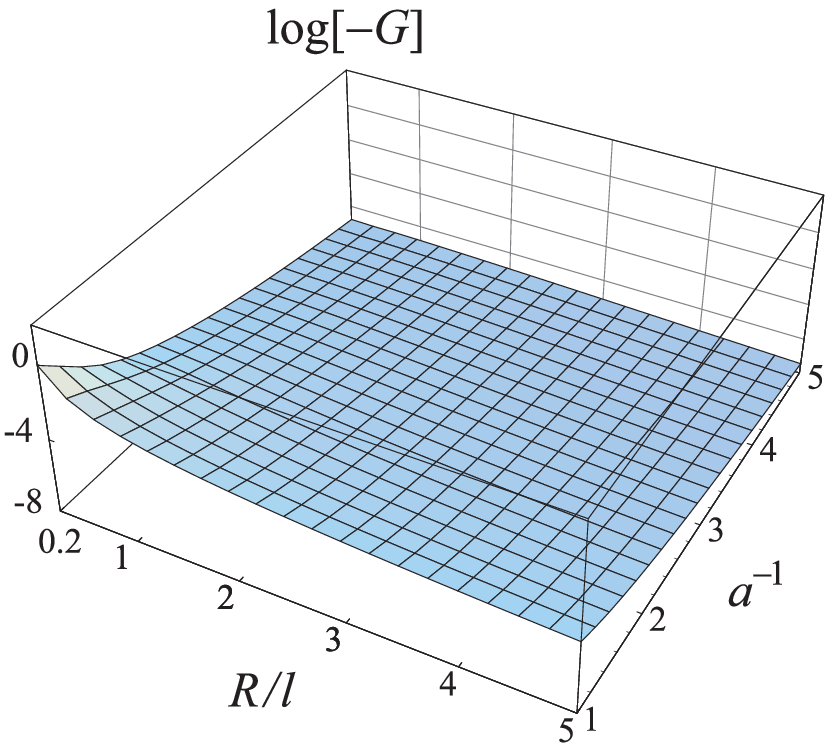}
  \epsfxsize = 6cm \
  \epsfbox{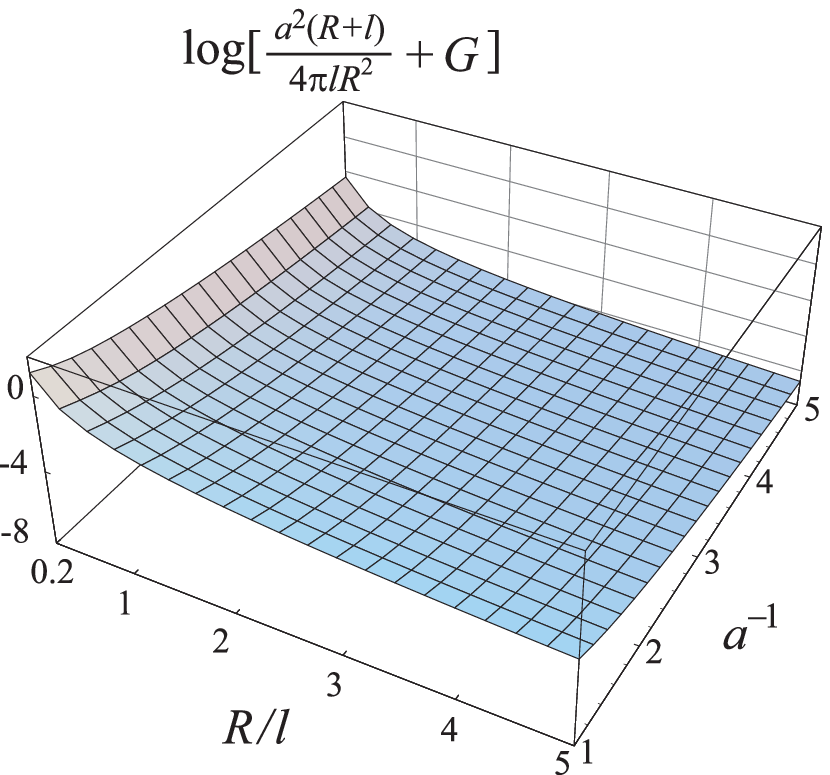}}
\end{center}
\caption{ 
We calculated numerically $G$ and $a^2(R+\ell)/(4\pi\ell R^2) +G $ to show the inequalities of (\ref{eq:inequality}). Since there appears no logarithmic divergence,  $G$ and $ a^2(R+\ell)/(4\pi\ell R^2) + G$ are positive definite at least in the region shown in the figures. 
}  
\label{fig:suppresion1}
\end{figure}

When we discuss second order perturbations later, 
we use the following inequalities 
\begin{eqnarray}
    0 \le  -G({\bf x},y;{\bf x}',0) \le \frac{a^2(y)}
                   {4\pi\ell R^2} (R+\ell) \,,  
    \label{eq:inequality}
\end{eqnarray}
This inequality is suggested by the asymptotic form of the Green function,  and it is confirmed by numerical calculations 
as shown in Fig.\ref{fig:suppresion1}. 

\subsection{linear perturbations}
\label{subsec:linear}

We start with linear perturbations. From Eq.(\ref{eq:formal sol. of sA^J}), ${\cal A}^{(1)}$ is given by 
\begin{equation}
   a^2{\cal A}^{(1)}={4 \kappa_5 \over 3}\int d^3x' 
     G({\bf x},y;{\bf x}',0) \rho^{(1)}(r'). 
\end{equation}
Suppose that $r_*$ is the radius of the star. 
When we consider the metric at a point outside the star 
$r-r_* \gg \ell$ or far from the 3-brane, 
we can safely replace 
$G$ by the approximation (\ref{asymptotic}) 
with the relative error of $O(\ell^2/r_*^2)$ 
\footnote{Here $r_*$ in the denominator is just inserted to adjust the 
dimensionality. It can be $r$ instead of $r_*$.}. 
Furthermore, at a field point far from the star, 
the matter distribution can be replaced with a point source. 
Then, we obtain  
$
    {\cal{A}}^{(1)} (r,y) \sim - 4\,G_4 M
    a(2a^2 r^2+3\ell^2)/3(a^2 r^2+ \ell^2)^{3/2} \,. 
$
By using Eq.~(\ref{eq:solved B^(J) in GN gauge}), we find 
$
    {\cal B}^{(1)} (r,y) \sim 4\,G_4 M a/3\sqrt{ a^2r^2+\ell^2 } \,. 
$

On the other hand, if we are interested in the metric 
induced on the brane, we can also set $y=0$. 
Even if we consider the interior of the star, 
the inequality (\ref{eq:inequality}) 
implies that the contribution to ${\cal A}^{(1)}$ from $G_K$ is, 
at most, of $O((\kappa_4 M_*/r_*) (\ell^2/r_*^2) \log(\ell/r_*))$, 
where $M_*\equiv 4\pi \int r^2 \rho(r) dr$.  
Hence, it is small by a factor of 
$O((\ell^2/r_*^2) \log(\ell/r_*))$ compared with 
the leading term. 
Then, neglecting the collections of this order or higher, 
we obtain 
\begin{eqnarray}
    {\cal{A}}^{(1)} (r,0) &\approx& {8\over 3}\phi,
    \\
    {\cal{B}}^{(1)} (r,0) &\approx& -{8\over 3r}\partial_r 
          \triangle^{-1}\phi\,,
\label{metric1sub}
\end{eqnarray}
where $\triangle ^{-1}$ is the inverse of the Laplacian operator, and 
we have introduced Newtonian potential 
\begin{equation}
  \phi \equiv 4\pi G_4 \triangle^{-1}\rho.
\label{eq:Newton potential}
\end{equation}

These are transformed to the isotropic GN gauge by using
(\ref{eq:gauge trans. - dA(J) dB(J) dC(J)}) with 
\begin{equation}
\stackrel{\scriptscriptstyle(1) \, }{\hat{\xi}^y} = -{\ell \phi\over 3}\,,
\quad
\stackrel{\scriptscriptstyle(1) \,}{\xi^r}
\approx  {2\over 3}\partial_r \triangle^{-1}\phi, 
\label{xi1sub}
\end{equation}
which are derived from 
(\ref{eq:xi^y^(J) and f^(J)})  
and (\ref{eq:isot gauge cond for xi^r 1}), respectively.  
Finally, the resulting metric functions turn out to be  
\begin{equation}
 -\bar{\cal A}{}^{(1)}(r,0)\approx \bar{\cal B}{}^{(1)}(r,0)
  =\bar{\cal C}{}^{(1)}(r,0)\approx -2\phi(r),  
\end{equation}
which agree with the result for the 4-dimensional Einstein gravity.

\section{second order perturbations}
\label{sec:second} 

For later convenience, we quote the result obtained 
in the preceding section as 
\begin{eqnarray}
    {\cal{A}}^{(2)}(r,0)  \equiv  
    A_{\Sigma} + A_{S} + A_Q
    - \frac{2}{\ell} \, \triangle ^{-1}  \int_0^{\infty} 
               dy' a^2  \, Q_{yy}  \,, 
\label{eq:definition of each term of A^(2)_tt}
\end{eqnarray}
where 
\begin{eqnarray}
    A_{\Sigma} &=& 2 {\kappa_5} \int dx'^3 G({\bf x},0;{\bf x}',0) 
       \Sigma ^{(2)}\, , \cr 
    A_S &=& 
      2 \int d^3x'\int_{0}^{\infty} dy' \, G ({\bf x},0;{\bf x}',y') \,
       {1\over {r'}^{2}} \partial_{r'} 
       \left(r'{}^2 {\cal{A}}^{(1)}_{,r'}(r',y') 
             {\cal{B}}^{(1)}(r',y')\right) \, , \cr 
    A_Q &=& \frac{4}{\ell}   \int d^3x' \int_{0}^{\infty} dy'
    \left[   \int^{y'}_{\infty}  
         G({\bf x},0;{\bf x}',y'') \, dy''\right] 
         a^2(y') Q_{yy}(r',y') \, .
\label{eq:split up A^(2)_tt into each sorce term}
\end{eqnarray}
The source term $\Sigma^{(2)}$ for $A_{\Sigma}$ 
is given in Eq.(\ref{sigma}). The third term in  
(\ref{eq:formal sol. of sA^J}), is separated into two pieces, $A_Q$ and the last term in (\ref{eq:definition of each term of A^(2)_tt}), by  
performing integration by parts. To obtain the expression 
of this last term, we have also used 
$\int d^3x'\int_0^\infty G\, dy'=\int d^3x'\int_0^\infty G_0\, dy'=(1/2)\triangle^{-1}$. 
Here the first equality follows from 
Eq.~(\ref{eq:orthogonality of u_m and 1}).

Since we are interested in the metric induced on the 3-brane, 
we also write down the expression for ${\bar{\cal{A}}}^{(2)}$ at $y=0$. 
Combining (\ref{eq:solved xi^(J)}), 
(\ref{eq:gauge trans. - dA(J) dB(J) dC(J)}), 
(\ref{eq:xi^y^(J) and f^(J)}) and 
(\ref{eq:definition of each term of A^(2)_tt}), we obtain 
\begin{eqnarray}
 {\bar{\cal{A}}}^{(2)} (r,0) = A_{\Sigma} +A_S +A_Q 
  + \frac{\kappa_4}{3}\triangle^{-1}T^{(2)} 
 + \Bigl[ ~\frac{2}{\ell} \triangle^{-1}~ \Xi -\frac{1}{2} 
   \Bigl( \stackrel{\scriptscriptstyle(1)\ }{ {\xi}^y_{,r}} \Bigr)^2 -  
   {\bar{\cal{A}}}^{(1)} _{,y}\!
   \stackrel{\scriptscriptstyle(1)\ }{ \xi^y} - \, {\bar{\cal{A}}}^{(1)} _{, r}
   \!\stackrel{\scriptscriptstyle(1)\ }{\xi^r}  \Bigr]_{y=0}   \,.
\end{eqnarray}
Here we note that 
the last term in (\ref{eq:definition of each term of A^(2)_tt})
is canceled with a term which arises from the gauge transformation.

The terms in the square brackets are evaluated just by substituting 
the estimate at the first order, and the first term $A_{\Sigma}$ 
has structure similar to first order perturbations, 
i.e., the source term is localized on the brane. 
Thus, the evaluation of these terms is straightforward. 
What needs detailed investigation  
is the evaluation of $A_S$ and $A_Q$. 
Since the source terms of $A_S$ and $A_Q$, which are 
quadratic in the linear perturbation quantities, 
distribute through the 5-dimensional bulk, 
it is necessary to evaluate a convolution of three 
5-dimensional Green functions.

Deferring the estimate of this convolution until
\S~\ref{subsec:suppression}, let us turn to the spatial components of
the second order metric perturbations. 
Although each spatial component 
depends on the choice of $\stackrel{\scriptscriptstyle(2)~}{\xi^r}$, 
the gauge invariant combination
\begin{eqnarray}
\left\{{\bar{\cal{B}} }^{(2)} - \partial_r(r{\bar{\cal{C}} }^{(2)})
\right\}_{y=0}
&=& \left\{{\cal{B}}^{(2)} - \partial_r(r{\cal{C}}^{(2)})
 -\delta {\cal{B}}^{(2)} + \partial_r(r\delta{\cal{C}}^{(2)})
  \right\}_{y=0}
\nonumber
\\
&=&
\left\{\frac{r}{2} {\bar{\cal{A}}}^{(2)}_{,r}
+ \frac{r}{2}\partial_r \delta {\cal{A}}^{(2)} 
  -\delta {\cal{B}}^{(2)}+
  \partial_r (r \delta {\cal{C}}^{(2)})
 \right\}_{y=0}
 - {r} \int^{\infty}_{0} dy ~Q_{yr} 
\label{Binv}
\end{eqnarray}
does not contain $\stackrel{\scriptscriptstyle(2)~}{\xi^r}$. 
Here we have used Eqs.~(\ref{eq:definition of phi}) and 
(\ref{eq:solved B^(J) in GN gauge}) with $y=0$
to eliminate ${\cal{B}}$ and ${\cal{C}}$.  
Taking the isotropic gauge
${\bar{\cal{B}} }^{(2)}={\bar{\cal{C}} }^{(2)}$, 
the left hand side becomes $-r{\bar{\cal{B}} }^{(J)}_{,r}$. 
Integrating this equation with respect to $r$, 
we obtain the expression for 
${\bar{\cal{B}} }^{(2)}(={\bar{\cal{C}} }^{(2)})$. 
Here we note that the terms in the square brackets 
in Eq.(\ref{Binv}) contains $\stackrel{\scriptscriptstyle(2)~}
{{\hat\xi}{}^{y}}$, from which a term with $y$-integration,
$$
 -{3\over \ell}\int_0^{\infty} dy ~a^2 Q_{yy} , 
$$
arises. Combined with the last term with $y$-integration 
in Eq.~(\ref{Binv}), this term is reduced to the expression 
that does not contain $y$-integration. 
The detail of calculation is 
explained in Appendix A. Here we just quote the final result 
\begin{eqnarray}
{\bar{\cal{B}} }^{(2)} (r,0) &=& 
 -\frac{1}{2} {\bar{\cal{A}}}^{(2)} 
 + \triangle^{-1} \left( \frac{\kappa_4}{2}T^{(2)} 
 + \frac{3}{\ell} \Xi
 + \frac{1}{2} Q_{yy} -{3\over 16}({\cal{A}}^{(1)}_{,r})^2
 \right)   
 +B_\Sigma  
 \,, \quad  ({\mathrm{ at ~~}} y=0) \,,
\label{Bbar}
\end{eqnarray}
where 
\begin{eqnarray}
B_\Sigma 
 \equiv
 - \frac{3}{4} \Bigl( 
     \stackrel{\scriptscriptstyle(1)}{\hat{\xi}^y_{,r}} \Bigr)^2
  + \int dr \frac{1}{r}  
    \bigl( \stackrel{\scriptscriptstyle(1)}{\hat{\xi}^y_{,r}} \bigr)^2
  + \frac{\ell \kappa_4}{4} \triangle^{-1}
    \left[  (3  {{\cal A}}^{(1)}_{,y}+  {{\cal B}}^{(1)}_{,y})
    \Sigma^{(1)} 
   - \frac{1}{r^3} (3  {{\cal B}}^{(1)}_{,y}+ {{\cal A}}^{(1)}_{,y}) \int r^2 \Sigma^{(1)}  dr  \right]_{y=0} \,.
\nonumber 
\end{eqnarray}
We find that the expression is reduced to a closed form written 
solely in terms of the quantities on the 3-brane except for 
${\bar{\cal{A}}}^{(2)}(r,0)$. 

\subsection{Recovery of the 4-dimensional Einstein gravity}

In this subsection, we evaluate the metric induced on the 3-brane at the 
leading order in $\ell/r_*$, and show that the result 
completely agrees with the one predicted by the 4-dimensional 
Einstein gravity. 

As for $A_{\Sigma}$, since the source term is localized on the 3-brane,  
we can approximate it as 
\begin{equation}
  A_{\Sigma}\approx 2\kappa_4 \triangle^{-1} \Sigma^{(2)},  
\end{equation}
which is justified 
for the same reason explained in evaluating the induced metric at 
the linear order. 

As mentioned earlier, the point which needs careful analysis 
is the computation of $A_S$ and $A_Q$. To evaluate these terms
we need to evaluate a convolution of the 5-dimensional Green functions. 
Since $G$ is composed of $G_0$ and $G_K$, 
the contribution from these terms 
is decomposed into several pieces depending on which combination of 
three propagators is used. 
For example, there is a mode coupling in which 
a zero mode propagator $G_0$ propagates the second order source 
produced by a product of KK mode contributions. 
We denote this mode coupling as 
\begin{eqnarray}
    [K, K;0] \,,
\nonumber
\end{eqnarray}
and similar labels are assigned for 
the other mode couplings, too.  

Here in the present subsection, we simply neglect the pieces 
containing the KK propagator $G_K$. 
Namely, we just take into account 
$A_{S[0,0;0]}$ and $A_{Q[0,0;0]}$. 
The justification of this approximation is given in 
the next subsections, where 
we prove that the neglected 
pieces in $A_S$ and $A_Q$ containing $G_K$ are actually 
suppressed by a factor of $O((\ell^2/r_*^2) \log(r_*/\ell))$. 

As long as only the contributions from the zero mode are concerned, 
the first order metric functions are all constant in $y$.   
Since the source term of $A_Q$ contains differentiation 
of the first order metric functions with respect to $y$, 
we find that $A_{Q[0,0;0]}\approx 0$. 
As for $A_{S[0,0;0]}$, after the $y$-integration,  
we find that it is reduced to 
\begin{equation}
 A_S\approx
  \triangle^{-1} {1\over r^2}\partial_r \left(r^2 {\cal{A}}^{(1)}_{,r} 
       {\cal{B}}^{(1)}\right). 
\end{equation}

Now that the evaluation of ${\bar{\cal{A}}}^{(2)} (r,0)$ is straightforward. Substituting the evaluations of the first order quantities  presented in \S~\ref{subsec:linear}, we finally obtain   
\begin{equation}
 {\bar{\cal{A}}}^{(2)} (r,0)
 \approx 
  \kappa_4 \triangle^{-1}
           \left(\rho^{(2)} -2\phi \rho^{(1)}+ 3P\right), 
\label{eq:leading for A^2}
\end{equation}
and 
\begin{eqnarray}
    {\bar{\cal{B}}^{(2)} } (r,0)={\bar{\cal{C}}^{(2)} } (r,0)
    \approx - \triangle^{-1} 
    \left[
 \kappa_4 \left(\rho^{(2)} -2\phi \rho^{(1)} \right)+(\phi_{,r})^2
  \right].     
\label{eq:leading for B^2}
\end{eqnarray}
It must be noted that $B_\Sigma$ in (\ref{Bbar}) gives only higher order correction of $O(\ell^2 \kappa_4^2)$ as it is easily shown by using (\ref{eq:Junction condition -explicit form}),  (\ref{eq:gauge trans. - dA(J) dB(J) dC(J)}), (\ref{eq:xi^y^(J) and f^(J)}) and  (\ref{eq:isot gauge cond for xi^r 1}) These results agree with those for the 4-dimensional Einstein 
gravity, whose brief derivation is given in Appendix B.

\subsection{Suppression of the KK-mode propagation}
\label{subsec:suppression}

In the preceding subsection the terms in $A_S$ and $A_Q$ 
containing a KK-mode propagator were neglected.  
Here we show that the contribution from these terms is in fact negligible. 
We begin with discussing rather general things. 
As for $A_S$, the terms $A_{S[0,0;K]}$, $A_{S[0,K;0]}$ and 
$A_{S[K,0;0]}$ vanish because of the 
orthogonality (\ref{eq:orthogonality of u_m and 1}). 
Hence the terms to be investigated are 
1.) $A_{S[K,K;0]}$, $A_{S[0,K;K]}$ $A_{S[K,0;K]}$ and $A_{S[K,K;K]}$. 
As for $A_Q$, the situation is a little simpler. 
Recall the fact mentioned above that the zero mode contribution 
in the first order metric functions is $y$-independent.  
Thus we can say that $A_{Q[0,*;*]}$ and $A_{Q[*,0;*]}$ vanish. 
Therefore, all the terms that we need to consider for $A_Q$ are 
2.) $A_{Q[K,K;0]}$ and $A_{Q[K,K;K]}$. 

\subsubsection{$A_{S[K,K;0]}$, $A_{S[K,0;K]}$ and $A_{S[0,K;K]}$}
\label{sec:A^2 [A,_r B]}
The source term of $A_{S}$ is rewritten as
\begin{eqnarray}
{1\over r^2}\partial_r \left(r^2 {\cal A}^{(1)}_{,r}
           {\cal B}^{(1)}\right)
 &=&{1\over r_*^2}
     \left[\left\{\left(r_*^2 \partial_r^2 
          -{r_*^2\over r} \partial_r\right){\cal A}^{(1)}\right\}
          \left(-{1\over r^3}\int^r dr r^2 {\cal A}^{(1)}\right)
        -\left({r_*^2\over r}\partial_r {\cal A}^{(1)}\right) 
             {\cal A}^{(1)}\right]. 
\end{eqnarray}
Each source term has the form of
\begin{equation}
  \left({\cal O}^{[1]}_r {\cal A}^{(1)}\right)
  \left({\cal O}^{[2]}_r {\cal A}^{(1)}\right)
\end{equation}
with
$
  {\cal O}^{[1,2]}_r=1$, ${r_*^2\over r}\partial_r$, 
      $r_*^2\partial^2_r$  
or $r^{-3}\int_0^r dr\, r^2$. 
By using the 
orthonormal condition 
(\ref{eq:orthogonality of u_m u_m}), 
we obtain the formula 
\begin{equation}
 \int_0^{\infty} dy\, a^{-2}\, G_K({\bf x}_1,y;{\bf x}_2,0)
   G_K({\bf x}_3,y;{\bf x}_4,0)
    ={1\over 2(4\pi)^2 R_{12} R_{34}}\mu(R_{12}+R_{34}),
\label{tatamikomi}
\end{equation}
where $R_{AB}\equiv |{\bf x}_A-{\bf x}_B|$ and 
\begin{equation}
    \mu(s) \equiv \int^{\infty}_{0} dm \ u_m(0)^2 e^{-m s}\,. 
\end{equation}
The function $\mu(s)$ is bounded by 
\begin{eqnarray}
  0 \le  \mu(s) < \frac{\ell}{2s(s+\ell)}. 
\label{boundmu}
\end{eqnarray}

We begin with $A_{S[K,K;0]}+A_{S[K,K;K]}$. We can derive an upper 
bound for the absolute value of each term as 
\begin{eqnarray}
 && \left\vert~ {2\over r_*^2}
       \int d^3x' \int_0^{\infty} dy'\, G({\bf x},0;{\bf x}',y')\, 
   \left[{\cal O}_{r_1}^{[1]}{\cal A}_K^{(1)}(r_1,y')
       {\cal O}_{r_3}^{[2]} {\cal A}_K^{(1)}(r_3,y')
        \right]_{r_1=r_3=r'}\right\vert\cr 
 &&\quad < {1\over 8\pi\ell r_*^2} 
    \int d^3x' \int_0^{\infty} dy'\, {a^2(y')\over R^2}(R+\ell)\, 
      \Biggl[\left\{{\cal O}_{r_1}^{[1]}{\cal A}_K^{(1)}(r_1,y')
        +{\cal O}_{r_1}^{[2]}{\cal A}_K^{(1)}(r_1,y')\right\}^2
\cr &&\hspace*{6cm} +
          \left\{{\cal O}_{r_1}^{[1]}{\cal A}_K^{(1)}(r_1,y')
        -{\cal O}_{r_1}^{[2]}{\cal A}_K^{(1)}(r_1,y')\right\}^2
         \Biggr]_{r_1=r'}, 
\end{eqnarray}
where 
${\cal A}_K^{(1)}$ is a part of ${\cal A}^{(1)}$ 
propagated by the KK mode propagator, 
and we have used (\ref{eq:inequality}). 
In below, we show that the each term on the right hand side of 
the above inequality is at most $O(\ell^2\log(r_*/\ell))$.  
We express the respective term as 
\begin{equation}
 {1\over r_*^2}\int d^3x' {R+\ell\over R^2} 
   \left[{\cal O}_{r_1}^{[1]}
       {\cal O}_{r_3}^{[2]} v(r_1,r_3)
        \right]_{r_1=r_3=r'}, 
\label{standardform}
\end{equation}
with  
\begin{eqnarray}
 v(r_1,r_3)&\equiv& \ell^{-1}\int_0^{\infty}dy\, a^2 
    {\cal A}_K^{(1)}(r_1) {\cal A}_K^{(1)}(r_2)\cr
  &=& {\kappa_4^2\ell\over 2r_1 r_3}
     \int_0^{r_*} dr_2 \,r_2 \Sigma^{(1)}(r_2)
     \int_0^{r_*} dr_4 \,r_4 \Sigma^{(1)}(r_4)
     \int_{|r_1-r_2|}^{r_1+r_2} dR_{12}
     \int_{|r_3-r_4|}^{r_3+r_4} dR_{34}\,
      \mu(R_{12}+R_{34})\, ,
\label{vdef}
\end{eqnarray}
where 
$v(r_1,r_3)$ has been rewritten by using Eq.~(\ref{tatamikomi}). 
Introducing $U(s)\equiv -\int_0^s ds'\int_{\infty}^{s'} ds'' \mu(s'')$, 
\begin{eqnarray}
 v(r_1,r_3)  
  = -{\kappa_4^2\ell\over 2r_1 r_3}
     \int_0^{r_*} dr_2\, r_2 \Sigma^{(1)}(r_2)
     \int_0^{r_*} dr_4\, r_4 \Sigma^{(1)}(r_4)
     &\Biggl(&U({r_1+r_2}+{r_3+r_4})-
           U({|r_1-r_2|}+{r_3+r_4})
   \cr  &&-
           U({r_1+r_2}+{|r_3-r_4|})+
           U({|r_1-r_2|}+{|r_3-r_4|})\Biggr). 
\end{eqnarray}
By construction, $U(0)=0$. 
The inequality (\ref{boundmu}) indicates that 
\begin{equation}
   0 \leq U(s) \leq {1\over 2}\left[s\log\left(1+{\ell\over s}
     \right)+\ell\log\left(1+{s\over \ell}\right)\right]. 
\end{equation}
As long as $s\alt r_*$, we find that 
$U(s)$ is at most $O(\ell \log(r_*/\ell))$.  

Let us consider the action of once or twice 
differentiation with respect to $r_1$ on $U$. 
Since $r_1$ appears in the arguments of $U$ only in 
the combination of $r_1+r_2$ or $|r_1-r_2|$, 
the differentiation with respect to $r_1$ can 
be replaced with that to $r_2$. 
Then, integrating by parts, finally the differentiation 
can be moved so that it acts on the source term 
$r_2 \Sigma^{(1)}(r_2)$. As long as first or 
second derivative is concerned, differentiation of $U$ 
does not appear as the boundary term.  
The same thing is true for the pair of $r_3$ 
and $r_4$. Now, using the bound for $U$ obtained above,  
we can conclude that 
${\cal O}_{r_1}^{[1]} {\cal O}_{r_3}^{[2]} v(r_1,r_3)$ 
is at most 
$O(\ell^2\log (r_*/\ell))$ for $r_1, r_3 \alt r_*$.
When $r_1, r_3 \agt r_*$, the argument of $\mu$ cannot 
be small.  Then, we can use the bound $\mu(s)<\ell/2s^2$.  
Therefore, ${\cal O}_{r_1}^{[1]}
       {\cal O}_{r_3}^{[2]} v(r_1,r_3)$ is 
$O(\ell^2)$ for $r_1, r_3 \agt r_*$.

To conclude that 
$A_{S[K,K;0]}+A_{S[K,K;K]}$ 
is at most $O((\ell^2/r_*^2)\log(r_*/\ell))$,  
we have to examine whether 
$[{\cal O}_{r_1}^{[1]}
{\cal O}_{r_3}^{[2]} $ $v(r_1,r_3)]_{r_1=r_3=r}$ 
behaves well at $r\to \infty$ and at $r\to 0$ 
so that the operation of $\int d^3x'R^{-2}(R+\ell)$ 
is well-defined.

{}First we consider the large $r$ limit. 
Since $r_2$ and $r_4$ are bounded by the 
radius of the star, $\mu (R_{12}+R_{34})$ can be replaced with
$\mu(r_{1}+r_{3})\approx \ell/(r_1+r_3)^2$ for large $r_1$ and/or $r_3$, 
and hence $v(r_1,r_3)\approx \ell^2(\kappa_4 M_*)^2/r_1 r_3(r_1+r_3)^2$.  
The operation of 
${\cal O}^{[1]}_{r_1}$ and ${\cal O}^{[2]}_{r_3}$ 
does not make the fall-off worse. 
For small $r_1$, 
\begin{eqnarray}
v(r_1,r_3) &=& {\kappa_4^2\ell\over r_3}
     \int_0^{r_*} dr_2\, r_2\Sigma^{(1)}(r_2)
     \int_0^{r_*} dr_4\, r_4\Sigma^{(1)}(r_4)
     \int_{|r_3-r_4|}^{r_3+r_4} dR_{34}
      \left(\mu(r_{2}+R_{34})
            +{r_1^2\over 6} \mu''(r_{2}+R_{34})+O(r_1^4)\right)\cr
  &=& {\kappa_4^2\ell\over r_3}
     \int_0^{r_*} dr_2 \left[\left(
          1 +{r_1^2 \over 6}\partial_{r_2}^2+O(r_1^4)\right)
          r_2\Sigma^{(1)}(r_2)\right]
     \int_0^{r_*} dr_4\, r_4 \Sigma^{(1)}(r_4)
     \int_{|r_3-r_4|}^{r_3+r_4} dR_{34} \mu (r_{2}+R_{34}). 
\end{eqnarray}
Therefore, ${\cal O}^{[1]}_{r_1} v(r_1,r_3)$ is finite  
at $r_1\to 0$.

{}For $A_{S[K,0;K]}$, a similar expression is obtained as 
\begin{equation}
 A_{S[K,0;K]}={2\over r_*^2}\int dr'\,r'
   \left[{\cal O}_{r_3}^{[1]} w(r_1,r_2,r_3)\right]_{r_1=r,r_2=r_3=r'}  
       {\cal O}_{r'}^{[2]}{\cal A}_0^{(1)}(r'), 
\end{equation}
with  
\begin{eqnarray}
 w(r_1,r_2,r_3)&\equiv& r_2\int d\Omega_2 \int_0^{\infty}dy'\,  
    G_K({\bf x}_1,0;{\bf x}_2,y') {\cal A}_K^{(1)}(r_3,y')\cr
  &=& {\kappa_4 \ell\over 4r_1 r_3}
     \int_0^{r_*} dr_4\, r_4 \Sigma^{(1)}(r_4)
     \int_{|r_1-r_2|}^{r_1+r_2} dR_{12}
     \int_{|r_3-r_4|}^{r_3+r_4} dR_{34}\,
      \mu(R_{12}+R_{34}),  
\label{wdef}
\end{eqnarray}
and ${\cal A}_0^{(1)}$ represents the part of ${\cal A}^{(1)}$ 
propagated by $G_0$.  
In the same way, we can show that $A_{S[K,0;K]}$ is at most 
$O((\ell^2/r_*^2)\log(r_*/\ell))$.

\subsubsection{$A_{Q[K,K;0]}$ and $A_{Q[K,K;K]}$}

$A_{Q[K,K;0]}$ and $A_{Q[K,K;K]}$ are also bounded as 
\begin{eqnarray}
|A_{Q[K,K;0]}&+&A_{Q[K,K;K]}|\nonumber\\
 &< &
       {1\over 2\pi\ell}\int d^3 x'{R+\ell\over R^2} 
        \int_0^{\infty}dy \, a^4(y) Q_{yy}({\bf x},y)\cr 
 & \leq & {1\over 8\pi\ell}\int d^3 x'{R+\ell\over R^2} 
        \Biggl\{\left\vert {\cal A}^{(1)}_{,y}{\cal A}_K^{(1)}+
                   {\cal B}^{(1)}_{,y}{\cal B}_K^{(1)}+
                   2{\cal C}^{(1)}_{,y}{\cal C}_K^{(1)}
                    \right\vert_{y=0} \cr
 &&\quad\quad 
         + \left\vert\int_0^\infty dy\, \left( 
      (a^4 {\cal A}^{(1)}_{,y})_{,y}
      {\cal A}_K^{(1)}+
      (a^4 {\cal B}^{(1)}_{,y})_{,y}
      {\cal B}_K^{(1)}+
      2(a^4 {\cal C}^{(1)}_{,y})_{,y}
      {\cal C}_K^{(1)}\right)
                    \right\vert\Biggr\}.
\end{eqnarray}
The first term in the curly brackets is at most 
$O((\ell^3/r_*^4)\log(r_*/\ell))$. 
By using the relations obtained from 
Eq.~(\ref{eq:master eq of delA^(J) in RS - Sigma}), 
\begin{eqnarray}
      {1\over a^4}(a^4 {\cal A}^{(1)}_{,y})_{,y}
     & = & -a^{-2}\triangle {\cal A}^{(1)}+2\kappa_5\Sigma^{(1)}\delta(y),
\label{nablaA}
\\
      {1\over a^4}(a^4 {\cal B}^{(1)}_{,y})_{,y}
     & = & a^{-2}{1\over r}\partial_r {\cal A}^{(1)} 
                 -{2\kappa_5\over r^3}\int dr \, r^2 \Sigma^{(1)}\delta(y),
\label{nablaB}
\end{eqnarray}
and ${\cal C}^{(1)}=-({\cal A}^{(1)}+{\cal B}^{(1)})/2$, 
the second term is reduced to the form in (\ref{standardform}). 
Therefore, the contribution to $A_Q$ from this term 
is at most $O((\ell^2/r_*^2)\log(r_*/\ell))$ for the same reason.

\section{summary}
\label{sec:summary}

In this paper we developed second order perturbations in the
RS single brane model, restricting the configuration to the static
axisymmetric one. From the 5-dimensional Einstein equations,
we derived the master equations for second order perturbations.  
At the level of linear perturbations, we can use the RS gauge, 
in which all perturbation equations are decoupled.  
Since the transverse-traceless condition cannot be imposed on
the second order, the second order perturbative equations are
inevitably coupled. 
As we have shown, however, the 4-dimensional spatial 
components of the second order metric function ${\cal{B}}^{(2)}$ and 
${\cal{C}}^{(2)}$ turned out to be concisely represented by the 
temporal component ${\cal A}^{(2)}$ with 
the first order metric functions. Therefore, 
the problem was reduced to solving a single differential equation 
for ${\cal{A}}^{(2)}$. 
Once we solve for ${\cal{A}}^{(2)}$, the other metric functions 
${\cal{B}}^{(2)}$ and ${\cal{C}}^{(2)}$ follow from it. 
Further, to discuss the metric induced on the 3-brane, 
we introduced the GN gauge in which a 
hypersurface with constant 5th-coordinate 
coincides with the location of the 3-brane. 
we gave the second order gauge transformations
between the RS gauge and the GN gauge explicitly.

Based on this formulation, we first discussed the zero mode
truncation for second order perturbations. It was shown that the
metric induced on the 3-brane evaluated by using the approximation of 
the zero mode truncation exactly
agrees with that for the 4-dimensional Einstein gravity.

Next, we evaluated the contribution to the 
metric functions from the KK modes. Since
the mode-by-mode analysis shows a pathological feature 
even at the level of linear perturbations, it is necessary to 
sum up all the mass eigenvalues to handle
the KK mode interactions. 
Performing such an analysis, we have
confirmed that the correction due to the KK mode coupling on the
induced metric is suppressed by a factor of 
$O((\ell^2/r_*^2) \log(r_*/\ell))$, and there
appears no pathological behavior. 
We therefore conclude that second order perturbations in the RS single brane model behave well and the result basically agrees with the prediction by the 4-dimensional Einstein gravity. 
The relative order of the correction is 
$O((\ell^2/r_*^2)\log(r_*/\ell))$. 
In the language of the post-Newtonian (PN) analysis of the 4-dimensional
Einstein gravity, the order of ${\cal A}^{(2)}$ and ${\cal B}^{(1)}$ is  1PN, and that of ${\cal B}^{(2)}$ is 2PN. According to the parameterized post-Newtonian (PPN) formalism, the PPN parameters at the 1PN order, $\beta$ and $\gamma$, are observationally constrained to the accuracy of about 0.1\% or so\cite{Will:1993}.  
Although this accuracy might be improved in the future, 
there will be no chance, unfortunately, to find the evidence 
for the large extra-dimension from the precision test of the 
solar system since the predicted 
deviation in $\beta, \,\gamma$ is $O((\ell^2/r_*^2)\log(r_*/\ell))$.

Of course, since our discussion developed in the present paper is 
restricted to second order perturbations with spherical symmetry,  
the gravity induced on the 3-brane in the RS single brane model 
might deviate from the 4-dimensional Einstein gravity in more 
general situation. Hence, it would be necessary to develop 
a more complete proof of the coincidence including all higher 
order terms would. Although we did not discuss the RS two brane model, it is also interesting to study the post Newtonian correction in this model. The analysis is now in progress and will be reported soon in a 
separate paper.

\centerline{\bf Acknowledgements}
H.K. would like to thank Takashi Nakamura and Hideo Kodama for informative comments and discussion.
We also like to thank V.F. Mukhanov for informing us Ref.\cite{Giannakis:2001zx}. 
This work is supported by the Monbukagakusho Grant-in-Aid No.~1270154.

\appendix

\section{Derivation of $\bar {\cal B}^{(2)}$}

As was mentioned in the text below Eq.~(\ref{Binv}), 
the expression given in the last line of Eq.~(\ref{Binv}) 
contain the terms with $y$-integration. 
They are explicitly written as 
\begin{eqnarray}
  \triangle^{-1} 
  \Biggl( - {3\over \ell} \int_0^{\infty} dy \, a^2 Q_{yy} 
       + {1\over r^2} \partial_r \int_0^\infty dy ~ r^2 Q_{yr}
  \Biggr). 
\end{eqnarray}
The first integral comes from 
$\stackrel{\scriptscriptstyle(2)}{\hat\xi^y}$ contained 
in $\delta{\cal A}^{(2)}$, $\delta{\cal B}^{(2)}$ and $\delta{\cal C}^{(2)}$. 
In obtaining the expression (\ref{Bbar}) from (\ref{Binv}), 
how to reduce these terms will be the only 
non-trivial manipulation. The rest of computation 
is slightly complicated but almost straightforward.

To rewrite the first integral, we use the relation 
\begin{eqnarray*}
 - \frac{1}{2} Q_{yy}(r,0) &=&{1\over 8}\int_0^{\infty}dy\,\partial_y \left\{a^{-6} 
          [(a^4{\cal A}^{(1)}_{,y})^2+(a^4{\cal B}^{(1)}_{,y})^2
           +2(a^4{\cal C}^{(1)}_{,y})^2]\right\} ,
\\
&=& {3 \over \ell} 
        \int_0^{\infty}dy\,a^{2}  Q_{yy}
  + {1\over 8} \int_0^\infty dy \frac{1}{a^2} 
    [(3{\cal A}^{(1)}_{,y}+{\cal B}^{(1)}_{,y})
           (a^4 {\cal A}_{,y}^{(1)})_{,y}
  + (3{\cal B}^{(1)}_{,y}+{\cal A}^{(1)}_{,y})
       (a^4 {\cal B}_{,y}^{(1)})_{,y} ],
\end{eqnarray*}
where we have used the traceless condition 
${\cal C}^{(1)}=-({\cal A}^{(1)}+{\cal B}^{(1)})/2$. 
On the other hand, the second part is rewritten as 
\begin{eqnarray*}
{1 \over r^2} \partial_r \int_0^\infty dy \, r^2 Q_{yr}      
  = - \frac{1}{2} Q_{rr}   
    + {1\over 8} \int_0^{\infty}dy \, 
        [(3{\cal A}^{(1)}_{,y}+{\cal B}^{(1)}_{,y})
           \triangle{\cal A}^{(1)}
           +(3{\cal B}^{(1)}_{,y}+{\cal A}^{(1)}_{,y})
           \triangle{\cal B}^{(1)}], 
\end{eqnarray*}
where again we have used the traceless condition.  
After the substitution of these relations and (\ref{nablaA}), the remaining $y$-integration 
is just 
\begin{equation}
 {1\over 8}\int_0^{\infty} dy\, 
         (3{\cal B}^{(1)}_{,y}+{\cal A}^{(1)}_{,y})
           \left[\triangle 
          + {1\over a^2} \partial_y a^4 \partial_y\right]
            {\cal B}^{(1)}. 
\label{remaining}
\end{equation}
We can derive
\begin{eqnarray}
      {1\over a^2}(a^4 {\cal B}^{(1)}_{,y})_{,y}
     & = & -\left(\triangle {\cal B}^{(1)}
                     +{2\over r} {\cal B}^{(1)}_{,r}\right)
                 -{2\kappa_5\over r^3}\int dr \, r^2 \Sigma^{(1)}\delta(y),
\end{eqnarray}
from Eq.~(\ref{nablaB}), 
and 
$(3{\cal B}^{(1)}_{,y}+{\cal A}^{(1)}_{,y})=-r {\cal B}^{(1)}_{,ry}$
from Eq.~(\ref{eq:solved B^(J) in GN gauge}). 
With the aid of these relations, it is easy to see that the remaining 
$y$-integration (\ref{remaining}) can be performed.

    \section{second order perturbations in the 4D Einstein Gravity}

In this Appendix, we give second order perturbations in the
4-dimensional Einstein gravity for the comparison with the zero mode
truncation of the gravity in the RS single brane model. Although 
the results are well known as the 2nd 
post-Newtonian analysis\cite{Chandrasekhar:1969}, 
we present a brief derivation for the following two reasons. 
1) Since we are working in a specific gauge, 
we need to consider a gauge transformation to compare our results 
with the expression presented in a different gauge. 
2) To compute metric perturbations in our restricted situation from 
the beginning is much easier than to follow a literature  
in which unrestricted cases are discussed.

We assume that the 4-dimensional metric
is static and isotropic, 
\begin{eqnarray*}
  ds^2 =-e^{  A  (r)} dt^2 + e^{ B (r)} d\textbf{x}^2 \,.
\end{eqnarray*}
Up to the second order, 
the 4-dimensional Einstein equations with the energy momentum tensor
given in (\ref{emtensor}) are 
\begin{eqnarray}
 R^t_t
&=&  \frac{1}{2}  (B-1) \triangle A
- \frac{1}{4} A_{,r} ( A +  B )_{,r}
= - \frac{\kappa_4}{2}(\rho+3P)  \,,
\label{eq:4D Ein tt}
\\
 R^r_r
&=& \frac{1}{2} (B-1) 
    \Bigl( \triangle ( A+2 B) 
   - \frac{2}{r} (A+B)_{,r } \Bigr) 
   + \frac{1}{4} A_{,r}(B-A)_{,r} 
   =  \frac{\kappa_4}{2} (\rho - P)  \,,
\label{eq:4D Ein rr}
\\
  R^\theta_\theta 
 &=& \frac{1}{2}(B-1) \Bigl( \triangle B 
  + \frac{1}{r}(A+ B)_{,r} \Bigr)  
  - \frac{1}{4} B_{,r}(A+B)_{,r}
  = \frac{\kappa_4}{2} (\rho - P) \,.
  \label{eq:4D Ein theta}
\end{eqnarray}
We expand $A$ and $B$ to the second order as (\ref{eq:def A^(J)}).  
Solving these equations at the linear order, we obtain $A ^{(1)}
=- B ^{(1)}  = 2\phi$, where $\phi(r)$ is similarly defined 
as in (\ref{eq:Newton potential}).  Putting these results into
Eq.~(\ref{eq:4D Ein tt}), the equation for $A^{(2)}$ becomes  
\begin{eqnarray}
   \triangle A^{(2)}    
=  {\kappa_4}  (\rho^{(2)}+3P) - 4 \phi \triangle \phi \,,
\label{eq:A^2 in 4D}
\end{eqnarray}
which agrees with (\ref{eq:leading for A^2}). 
Eliminating  $\triangle A^{(2)}$ and $P$ 
from Eqs.~(\ref{eq:4D Ein tt}), (\ref{eq:4D Ein rr}) and 
(\ref{eq:4D Ein theta}), 
we obtain 
\begin{eqnarray}
  \triangle B^{(2)}    
=  - {\kappa_4} \rho^{(2)} + 4 \phi \triangle \phi
  - (\phi_{,r})^2 \,.
\label{eq:B^2 in 4D}
\end{eqnarray}
Hence it was shown that the result obtained by the zero mode truncation 
(\ref{eq:leading for B^2})
agrees with that for the 4-dimensional Einstein gravity (\ref{eq:B^2 in 4D}).

  	


\begin{thebibliography}{99}


\bibitem{Arkani-Hamed:1998rs}
N.~Arkani-Hamed, S.~Dimopoulos and G.~Dvali,
Phys.\ Lett.\ {\bf B429}, 263 (1998);
Phys.\ Rev.\ D {\bf 59}, 086004 (1999).

\bibitem{Antoniadis:1998ig}
I.~Antoniadis, N.~Arkani-Hamed, S.~Dimopoulos and G.~Dvali,
Phys.\ Lett.\ {\bf B436}, 257 (1998).

\bibitem{Randall:1999ee}   L.~Randall and R.~Sundrum,
Phys.\ Rev.\ Lett.\ {\bf 83}, 3370 (1999).

\bibitem{Randall:1999vf}    L.~Randall and R.~Sundrum,
Phys.\ Rev.\ Lett.\ {\bf 83}, 4690 (1999).
  
 

\bibitem{Garriga:2000yh}   J.~Garriga and T.~Tanaka,
Phys.\ Rev.\ Lett.\ {\bf 84}, 2778 (2000).


\bibitem{Shiromizu:2000wj}   T.~Shiromizu, K.~Maeda and M.~Sasaki,
Phys.\ Rev.\ D {\bf 62}, 024012 (2000).

\bibitem{Sasaki:2000mi}    M.~Sasaki, T.~Shiromizu and K.~Maeda,
Phys.\ Rev.\ D {\bf 62}, 024008 (2000).

\bibitem{Gen:2000nu}
U.~Gen and M.~Sasaki,
gr-qc/0011078.

\bibitem{Tanaka:2000er}
T.~Tanaka and X.~Montes,
Nucl.\ Phys.\ {\bf B582}, 259 (2000).


\bibitem{Giddings:2000mu} S.~B.~Giddings, E.~Katz and L.~Randall,
JHEP{\bf 0003}, 023 (2000).

\bibitem{Tanaka:2000zv}   T.~Tanaka,
Prog.\ Theor.\ Phys.\ {\bf 104}, 545 (2000).

 
\bibitem{Csaki:2000mp}
C.~Csaki, M.~Graesser, L.~Randall and J.~Terning,
Phys.\ Rev.\ D {\bf 62}, 045015 (2000).


\bibitem{Charmousis:2000rg}
C.~Charmousis, R.~Gregory and V.~A.~Rubakov,
Phys.\ Rev.\ D {\bf 62}, 067505 (2000).




\bibitem{Chiba:2000rr}
T.~Chiba,
Phys.\ Rev.\ D {\bf 62}, 021502 (2000).


\bibitem{Goldberger:1999uk}  W.~D.~Goldberger and M.~B.~Wise,
Phys.\ Rev.\ Lett.\ {\bf 83}, 4922 (1999).

\bibitem{Goldberger:2000un}   W.~D.~Goldberger and M.~B.~Wise,
Phys.\ Lett.\ {\bf B475}, 275 (2000).

\bibitem{DeWolfe:2000cp}
O.~DeWolfe, D.~Z.~Freedman, S.~S.~Gubser and A.~Karch,
Phys.\ Rev.\ D {\bf 62}, 046008 (2000).

 


\bibitem{Kodama:2000fa}
H.~Kodama, A.~Ishibashi and O.~Seto,
Phys.\ Rev.\ D {\bf 62}, 064022 (2000).


 
\bibitem{Mukohyama:2000ui}
S.~Mukohyama,
Phys.\ Rev.\ D {\bf 62}, 084015 (2000).
,
Class.\ Quant.\ Grav.\ {\bf 17}, 4777 (2000).



\bibitem{Maartens:2000fg}
R.~Maartens,
Phys.\ Rev.\ D {\bf 62}, 084023 (2000).

\bibitem{Koyama:2000cc}
K.~Koyama and J.~Soda,
Phys.\ Rev.\ D {\bf 62}, 123502 (2000).

\bibitem{vandeBruck:2000ju}
C.~van de Bruck, M.~Dorca, R.~H.~Brandenberger and A.~Lukas,
Phys.\ Rev.\ D {\bf 62} 123515 (2000). 

\bibitem{Langlois:2000ia}
D.~Langlois,
Phys.\ Rev.\ D {\bf 62}, 126012 (2000).

 


\bibitem{Chamblin:2000cj}
A.~Chamblin and G.~W.~Gibbons,
Phys.\ Rev.\ Lett.\ {\bf 84}, 1090 (2000).
 

\bibitem{Chamblin:2000by}
A.~Chamblin, S.~W.~Hawking and H.~S.~Reall,
Phys.\ Rev.\ D {\bf 61}, 065007 (2000).
 
 

\bibitem{Gregory:1993vy}
R.~Gregory and R.~Laflamme,
Phys.\ Rev.\ Lett.\ {\bf 70} 2837 (1993). 



\bibitem{Gregory:2000gf}
R.~Gregory,
Class.\ Quant.\ Grav.\ {\bf 17}, L125 (2000).

\bibitem{Chamblin:2001ra}
A.~Chamblin, H.~S.~Reall, H.~Shinkai and T.~Shiromizu,
Phys.\ Rev.\ D {\bf 63}, 064015 (2001).

\bibitem{Emparan:2000wa}
R.~Emparan, G.~T.~Horowitz and R.~C.~Myers,
JHEP{\bf 0001}, 007 (2000).

\bibitem{Dadhich:2000am}
N.~Dadhich, R.~Maartens, P.~Papadopoulos and V.~Rezania,
Phys.\ Lett.\ B {\bf 487}, 1 (2000).

 
\bibitem{Israel:1966}
W. Israel, Nuovo Cim. {\bf 44B}, 1 (1966).


\bibitem{Will:1993}
C.M. Will,  
{\it Theory and Experiment in Gravitatinal Physics}
(Cambridge University Press, Cambridge 1993), 
revised ed; Lecture notes from the 1998 Slac Summer Institute on 
Particle Physics, gr-qc/9811036. 

\bibitem{Chandrasekhar:1969}
e.g., S.~Chandrasekhar and Y. Nutku, 
Astrophy. J. {\bf 158}, 55 (1969);
L.~Blanchet,
Phys.\ Rev.\ D {\bf 51}, 2559 (1995). 

\bibitem{Giannakis:2001zx}
I.~Giannakis and H.~Ren,
Phys.\ Rev.\ D {\bf 63}, 024001 (2001)


    \end{thebibliography}
\end{document}